\documentstyle[twoside,fleqn,espcrc2]{article}

\def\beq{\begin{equation}}
\def\eeq{\end{equation}}

\def\td{\right)} 
\def\ts{\left(}

\def\bl{\beta^{\scriptscriptstyle {LAT}}}
\def\gl{\gamma^{\scriptscriptstyle {LAT}}}
\def\bms{\beta^{\scriptscriptstyle {\overline{MS}}}}
\def\gms{\gamma^{\scriptscriptstyle {\overline{MS}}}}
\def\ccin{c_5}
\def\csei{c_6}
\def\cset{c_7}
\def\cott{c_8}
\def\cnov{c_9}

\newcommand{\AmS}{{\protect\the\textfont2
  A\kern-.1667em\lower.5ex\hbox{M}\kern-.125emS}}

\title{Four loop results for the 2D $O(n)$ nonlinear $\sigma$ model
  with 0--loop and 1--loop Symanzik actions}

\author{B.~All\'es and M.~Pepe\thanks{Poster presented by M. Pepe.}\\
Dipartimento di Fisica, Universit\`a di Milano-Bicocca
and INFN, Sezione di Milano, Italy}
\begin{document}

\begin{abstract}
We present complete three loop results and preliminary four loop
results for the 2D  $O(n)$ nonlinear
$\sigma$ model with 0--loop and 1--loop Symanzik improved actions. 
This calculation 
aims to test the improvement in the numerical precision that the
combination of Symanzik actions and effective couplings
can give in Monte Carlo simulations.
\end{abstract}

\maketitle

\section{INTRODUCTION}

Monte Carlo simulations on the lattice are affected by systematic
errors due to the 
finiteness of the lattice spacing $a$. Symanzik proposed a method to
reduce these effects on the physical scaling by following a
perturbative procedure~\cite{Sym}.

The integration of the beta function of the theory allows
to express the results of a simulation in physical units by fixing the
lattice scale $a$. The asymptotic scaling regime is attained when 
the lattice scale is well determined with the first universal
terms in the perturbative expansion of the beta function. However this
regime is barely achieved and an estimate of the non--universal
corrections to asymptotic scaling is necessary.

The asymptotic scaling regime can be more easily accomplished, even for
moderately large couplings, if an effective coupling~\cite{Par} is
used. 

We plan to test the numerical precision obtained in 
Monte Carlo simulations by using a combination of Symanzik improved actions
together with an effective coupling. We want to perform this test on the 2D
$O(n)$ nonlinear $\sigma$ model. In this proceeding we present
the perturbative calculation of the Renormalization Group Invariant (RGI)
functions. We also report on a recent calculation of
the weak coupling expansion of the energy up to four loops~\cite{noi}
which is necessary to express the beta function in terms of the
effective coupling.

\section{SYMANZIK IMPROVED ACTIONS}

We consider the 0--loop and the 1--loop Symanzik improved actions: 
\begin{eqnarray}\label{az}
&&\hskip -8mm S^{\rm 0-loop} = {a^2 \over g} \sum_{x} 
                     \left( {2 \over 3} \vec\phi_x  K_1  \vec\phi_x -
                            {1 \over 24} \vec\phi_x  K_2  \vec\phi_x
                     \right)\;,  \nonumber \\ 
&&\hskip -8mm S^{\rm 1-loop} =   {a^2 \over g} \sum_x \bigg[ \;{1 \over 2}
         \vec\phi_x  K_1  \vec\phi_x -
         a^2\ccin \;g\left(K_1  \vec\phi_x\right)^2  \nonumber \\
&&\hskip -8mm    - a^2\left(\csei \; g\,- {1 \over 24}\right) 
        \sum_\mu \left( \partial^+_\mu\partial^-_\mu \vec\phi_x\right)^2 
      \nonumber \\
&&\hskip -8mm    - a^2\cset \;g\left(\vec\phi_x  K_1  
         \vec\phi_x\right)^2 - a^2\cott\;g
         \sum_\mu \left(\vec\phi_x 
         \partial^+_\mu\partial^-_\mu \vec\phi_x\right)^2 
  \nonumber \\
&&\hskip -8mm      -  {1 \over 16} a^2 \cnov \;g\sum_{\mu\nu} \left(
        \left(\partial^+_\mu + \partial^-_\mu\right) \vec\phi_x \cdot
        \left(\partial^+_\nu + \partial^-_\nu\right) \vec\phi_x \right)^2 \bigg]
      \;. \nonumber \\
&&
\end{eqnarray}
The $n$--component scalar field $\vec\phi _x$ is constrained by
$\vec\phi^2 _x =1$.  The lattice operators in (\ref{az}) are
\begin{eqnarray}
 K_1 \vec\phi_x & \equiv & {1\over a^2}\sum_\mu \left( 2 \vec\phi_x - 
               \vec\phi_{x+\hat\mu} - 
               \vec\phi_{x-\hat\mu} \right)\;, \nonumber \\
 K_2 \vec\phi_x & \equiv & {1\over a^2}\sum_\mu \left( 2 \vec\phi_x - 
               \vec\phi_{x+2\,\hat\mu} - 
               \vec\phi_{x-2\,\hat\mu} \right)\;, \nonumber \\
 \partial^+_\mu \vec\phi_x &\equiv& {1\over a}\left(\vec\phi_{x+\hat\mu} - 
             \vec\phi_x\right)\;,  \nonumber \\
 \partial^-_\mu \vec\phi_x &\equiv& {1\over a}\left(\vec\phi_x - 
             \vec\phi_{x-\hat\mu} \right)\;.
\end{eqnarray}

The $c_i$ coefficients are fixed by the Symanzik improvement program at 
one loop~\cite{Sym}.

\section{ASYMPTOTIC SCALING CORRECTIONS}

In order to know the corrections to asymptotic scaling up to four
loops we need to compute the perturbative expansion of
$\bl$ and $\gl$ up to four loops for the two considered Symanzik
actions. In this proceeding we report the results at three loops while
the evaluation for the next order is in progress.

The knowledge of the continuum
RGI functions $\bms$, $\gms$ up to  
four loops allows the calculation of the analogous lattice functions
$\bl$, $\gl$ up 
to four loops by a three--loop computation of the 2--point 
1PI correlation function $\Gamma^{(2)}_{LAT}$.

We have treated the constraint on the norm of the field $\vec\phi$
with the standard method~\cite{Zinn}. As a consequence the theory is
described in terms of a $(n-1)$--component 
field $\vec\pi$ and a measure term has to be added to the action. 

The expressions relating the RGI functions
on the lattice and in the $\overline{MS}$ scheme are
\begin{eqnarray}
\bl (g) = {{Z^g  (g) \bms(g_R) } \over {\displaystyle 1 - g_R 
{\strut\partial Z^g (g) \over \displaystyle \partial g}}} \;,
 \qquad\qquad \qquad\qquad \nonumber \\
\gl (g) = \gms (g_R) - \bl (g) {\partial \log Z^\pi (g) \over \partial g}\;,\;\;\;
\end{eqnarray}
where $Z^g$ and $Z^\pi$ are the
renormalization constants of the coupling and the field respectively;
$g$ and $g_R$ are the bare and the renormalized coupling constants.

We write the perturbative expansions as follows:
$\bl=-\beta_0 \; g^2 -\beta_1 \;g^3 -\beta^{\scriptscriptstyle {LAT}}_2 \;g^4 
-\ldots $ and $\gl=\gamma_0 \; g +\gamma^{\scriptscriptstyle {LAT}}_1 \;g^2 
+\ldots$ The coefficients $\beta_0$, $\beta_1$, $\gamma_0$ are universal; 
our result at three loops for the 0--loop action is
\begin{eqnarray}\label{rg0}
&& \hskip -.6cm \beta_{2,0-loop}^{\scriptscriptstyle {LAT}}=\frac{(n -2 )}{16\pi}
\Big[ (n -2 ) \ts -1 +\frac{1}{\pi^2} -8 G_1^S 
\Big. \right.  
\nonumber \\
&& \hskip -.5cm \Big. \left. 
+\frac{5}{6} Y_1-\frac{7}{48} Y_1^2 -\frac{2}{3} Y_{1,2}+\frac{5}{18} Y_1 Y_{1,2} \td-
1+\frac{2}{\pi^2} 
\Big.  
\nonumber \\
&& \hskip -.5cm \Big. 
-\frac{4}{27} G_2^S+\frac{5}{6} Y_1+\frac{1}{3 \pi} Y_1
-\frac{29}{144} Y_1^2-\frac{1}{18} Y_2 \Big.
\nonumber \\
&& \hskip -.5cm \Big. +\frac{5}{216} Y_1 Y_2 \Big] \;,
\nonumber \\
&& \hskip -.5cm \gamma_{1,0-loop}^{\scriptscriptstyle
{LAT}}=\frac{(n-1)}{24\pi} Y_1  \;,
\nonumber \\
&& \hskip -.5cm \gamma_{2,0-loop}^{\scriptscriptstyle {LAT}}=\frac{(n-1)}{16\pi}
\Big[(n -2 ) \ts 1 +\frac{1}{\pi^2}+8 G_1^S 
\Big. \right. 
\nonumber \\
&& \hskip -.4cm \Big. \left. 
-\frac{5}{6} Y_1+\frac{7}{48} Y_1^2 -\frac{2}{3} Y_{1,2} 
+\frac{5}{18} Y_1 Y_{1,2} \td 
\Big.  
\nonumber \\
&& \hskip -.4cm \Big. 
+1+\frac{4}{27} G_2^S-\frac{5}{6} Y_1+\frac{37}{144} Y_1^2 \nonumber \\
&&\hskip-0.4cm+\frac{1}{18} Y_2 -\frac{5}{216} Y_1 Y_2 \Big] \;,
\label{0loop}
\end{eqnarray}
the notation being
\begin{eqnarray}\label{defY}
&&  \hskip -.5cm Y_{i,j} \equiv \int_{-\pi}^{+\pi} {\hbox{d}^2 q \over \left(2\pi\right)^2}
  {\left(\Box_q\right)^i \over \left(\Pi_q\right)^j} \;,
  \qquad\qquad Y_i \equiv Y_{i,i} \;,\nonumber \\
&& \hskip -.5cm  G_1^S \equiv -\int_{-\pi}^{+\pi} {\cal{D}}_3\;
\frac
{\sum_\mu \left( {1\over 18} \hat{l}^6_\mu + {1\over 144}
              \hat{l}^8_\mu \right) \Delta^S_{q,k}}
{\Pi_q\Pi_k\left(\Pi_l\right)^2}\;,
 \nonumber \\
&& \hskip -.5cm  G_2^S \equiv \int_{-\pi}^{+\pi} {\cal{D}}_3\;
{{3\; {\hat{l}^4_\mu}\; {\hat{q}^4_\mu}\; {\hat{k}_\mu}^2 - {1\over 4}\;
        \hat{l}^4_\mu\; \hat{q}^4_\mu\; \hat{k}^4_\mu } \over
        {\Pi_q\Pi_k\Pi_l}} \;,
\label{integrals}
\end{eqnarray}
where we use the standard notation 
$\hat{q}_{\mu} \equiv 2 \sin (q_\mu/2)$ and
\begin{eqnarray}
&&\hskip -5mm \hat{q}^2\equiv \sum_\mu \hat{q}^2_\mu \;,\qquad\qquad
 \Box_q\equiv \sum_\mu \hat{q}^4_\mu \;,\nonumber \\
&&\hskip -5mm \Pi_q\equiv \hat{q}^2 + {1\over 12} \Box_q\;, \;\;
 \Delta^S_{q,k} \equiv\Pi_{q+k} -\Pi_{q} -\Pi_{k}\;.
\end{eqnarray}
The measure term in the two--loop integrals is
\begin{equation}
 {\cal{D}}_3 \equiv 
 {\hbox{d}^2q \over \left(2\pi\right)^2}\,
 {\hbox{d}^2k \over \left(2\pi\right)^2}\,
 {\hbox{d}^2l \over \left(2\pi\right)^2}\;
 \left(2\pi\right)^2\,\delta(q+k+l) \;.
\end{equation}

Eq.(\ref{0loop}) in numerical form is
\begin{eqnarray}
&&\hskip -7mm\beta_{2,0-loop}^{\scriptscriptstyle {LAT}}=\frac{(n -2 )}{(2\pi)^3}
\Big[ 0.481294 + 0.181889 \; ( n -2 ) \Big]\;,  \nonumber \\
&&\hskip -7mm\gamma_{1,0-loop}^{\scriptscriptstyle {LAT}}=\frac{1}{(2\pi)^2}
\Big[ 1.07001 \; (n - 1 ) \Big]\;, \nonumber \\
&&\hskip -7mm\gamma_{2,0-loop}^{\scriptscriptstyle {LAT}}=\frac{(n-1)}{(2\pi)^3}
\Big[ 2.73365 + 0.355965 (n-2)\Big]\;.
\end{eqnarray}
The agreement with Ref.~\cite{FalTre} is satisfactory within the
precision of the numerics.
The analogous coefficients for the 1--loop action are a new result,
\begin{eqnarray}\label{rg1}
&& \hskip -.5cm 
\beta_{2,1-loop}^{\scriptscriptstyle {LAT}}=\beta_{2,0-loop}^{\scriptscriptstyle{LAT}}+
\frac{(n - 2)}{2 \pi} (\eta -2 \zeta) \;,
\nonumber \\
&& \hskip -.5cm 
\gamma_{1,1-loop}^{\scriptscriptstyle {LAT}}= \gamma_{1,0-loop}^{\scriptscriptstyle {LAT}}\;,
\nonumber \\
&& \hskip -.5cm 
\gamma_{2,1-loop}^{\scriptscriptstyle {LAT}}=\gamma_{2,0-loop}^{\scriptscriptstyle {LAT}}
-\frac{(n - 1)}{2 \pi} \eta -\frac{(n - 3)}{\pi} \zeta\;,
 \nonumber \\
&& \hskip -.5cm \eta\equiv
 (n - 1) \ts c_8 (\frac{1}{6} Y_1 - 2) + c_7 (\frac{1}{3} Y_1 -4)
\right. \nonumber \\
&& \hskip -.4cm \left. + c_9 (\frac{2}{3} Y_1 - 2)\td
+(\frac{4}{3} Y_1 - 4) (c_7 + c_8 + \frac{3}{2} c_9) 
\nonumber \\
&& \hskip -.4cm + c_6 (\frac{5}{2} Y_1 + 2 Y_{1,2} - \frac{1}{6} Y_2 - 6) +
  c_5 (\frac{4}{3} Y_1 + \frac{1}{72} Y_2 
\nonumber \\
&& \hskip -.4cm +\frac{5}{144} Y_{2,1} - \frac{1}{864} Y_{3,2} - 5)  \;,
\nonumber \\
&&\hskip -.5cm \zeta\equiv 
(n-1)\ts c_6 Y_{1,2} + c_5 (1 - \frac{1}{6} Y_1 + \frac{1}{144} Y_2)\td \;.
\end{eqnarray}
 
\section{EFFECTIVE SCHEME}
The energy operator $E$ for the two improved actions is (no summation
over $\mu$)
\begin{equation}
 E= \langle {4\over 3}\vec\phi_0 
            \cdot \vec\phi_{0+\hat{\mu}}
            -{1\over 12} \vec\phi_0 \cdot \vec\phi_{0+2\hat{\mu}}\rangle \;.
\end{equation}
We write its perturbative expansion as 
$E=w_0 -w_1\; g -w_2\;  g^2 -\ldots$.
An effective scheme~\cite{Par} is introduced by defining the effective 
coupling constant $g_E$ 
\begin{equation}
 g_E \equiv {w_0 -E^{\rm MC} \over w_1}\;,
\end{equation}
where $E^{\rm MC}$ is the Monte Carlo measured value of $E$ at the
bare coupling $g$.
In order to express the asymptotic scaling corrections in terms of
$g_E$, we have calculated the
perturbative expansion of $E$ up to four loops for the two Symanzik
improved actions. We have first
put the model into a square box of finite size $L$ with periodic
boundary conditions in order to regularize the IR divergences. This
procedure has three consequences~\cite{Has}:\\
{\it{i}}) the momenta are summed, not integrated,\\
{\it{ii}}) the zero modes are absent,\\
{\it{iii}}) a new term, coming from a Faddeev--Popov
determinant, has to be added to the action.

The result for $E$ must be finite after the limit
$L\rightarrow\infty$ has been taken. However partial contributions from
individual diagrams can diverge in the thermodynamic limit. These divergent
contributions have been algebraically worked out to separate their
finite part from the divergent one. At the end all divergences cancel
out leaving a result that allows the limit $L\rightarrow\infty$.
After this limit, the sums over momenta become integrals in the
Brillouin zone.

The above--described calculation requires the evaluation of
diagrams containing vertices from the Faddeev--Popov
determinant~\cite{Has}. We have checked that the whole contribution of
these diagrams up to four loops 
vanishes in the limit $L\rightarrow\infty$.

More details as well as the analytical and the numerical results 
for the perturbative expansion of $E$ can be found in \cite{noi}.

\section{NUMERICAL TECHNICALITIES}
We have used three methods to compute the finite lattice
integrals in (\ref{integrals}), (see~\cite{noi})\\
{\it i}) extrapolation to the infinite lattice size
of the result on finite lattices,\\
{\it ii}) Gauss method, \\
{\it iii}) the coordinate space method~\cite{lusweisz} extended to the
case of improved propagators.\\
The results of our integrals are
\begin{eqnarray}
&&  \hskip -.5 cm  Y_1=2.0435764382979844236\;,  \nonumber \\
&&  \hskip -.5 cm  Y_2=4.7830710733439886212\;, \nonumber \\
&&  \hskip -.5 cm  Y_{1,2}=0.4729502261432961899\;, \nonumber\\ 
&&  \hskip -.5 cm  Y_{2,1}=30.077096804291341057\;, \nonumber \\
&&  \hskip -.5 cm  Y_{3,2}=77.324121011413132160\;, \nonumber \\
&&  \hskip -.5 cm  G_1^S=0.013948510\;, \nonumber \\
&&  \hskip -.5 cm  G_2^S=0.9748227\;.
\end{eqnarray}


\begin{thebibliography}{9}
\bibitem{Sym} K. Symanzik, Nucl. Phys. B226 (1983) 205.
\bibitem{Par} G. Parisi in Proc. XXth Int. Conf. on High Energy
Physics, (Madison, WI, 1980), edited by L. Durand and L. G. Pondrom;
G. Martinelli, G. Parisi and R. Petronzio, Phys. Lett. B100 (1981) 485.
\bibitem{noi} B. All\'es and M. Pepe, hep--lat/9906014.
\bibitem{Zinn} E. Br\'ezin, J. Zinn--Justin and J. C. Le Guillou,
Phys. Rev. D14 (1976) 2615.
\bibitem{FalTre} M. Falcioni and A. Treves, Nucl. Phys. B265 (1986) 671.
\bibitem{Has} P. Hasenfratz, Phys, Lett. B141 (1984) 385.
\bibitem{lusweisz} M. L\"uscher and P. Weisz, Nucl. Phys. B445 (1995) 429.
\end{thebibliography}
\end{document}